\documentclass{article}
\usepackage{amssymb}

\usepackage{graphicx}
\usepackage{amsmath}

\newtheorem{theorem}{Theorem}
\newtheorem{acknowledgement}[theorem]{Acknowledgement}

\newtheorem{definition}[theorem]{Definition}

\newtheorem{lemma}[theorem]{Lemma}

\newenvironment{proof}[1][Proof]{\textbf{#1.} }{\ \rule{0.5em}{0.5em}}
\input{tcilatex}

\begin{document}

\title{The Chebotarev-Gregoratti Hamiltonian as singular perturbation of a
nonsemibounded operator}
\author{John~Gough\thanks{%
Institute for Mathematics and Physics, University of Wales, Aberystwyth,
Ceredigion, SY23 3BZ, Wales}}
\date{}
\maketitle

\begin{abstract}
We derive the Hamiltonian associated to a quantum stochastic flow by
extending the Albeverio-Kurasov construction of self-adjoint extensions to
finite rank perturbations of nonsemibounded operators to Fock space.
\end{abstract}

\section{Introduction}

To analyze a dynamical system with Hamiltonian $K=K_{0}+\Upsilon $,
occurring as a perturbation of a free Hamiltonian $K_{0}$, we transfer to
the interaction picture by means of the wave operators 
\begin{equation*}
V\left( t\right) =U_{0}\left( -t\right) U\left( t\right)
\end{equation*}
where $U$ and $U_{0}$ are the strongly continuous groups generated by $K$
and $K_{0}$ respectively. The family $V=\{V\left( t\right) :t\geq 0\}$ then
satisfies the differential equation 
\begin{equation*}
\dot{V}\left( t\right) =-i\Upsilon \left( t\right) V\left( t\right)
\end{equation*}
where $\Upsilon \left( t\right) =U_{0}\left( -t\right) \Upsilon U_{0}\left(
t\right) $. However, $V$ is not a semi-group, but instead forms a $U_{0}$%
-cocycle, that is $V\left( t+s\right) =U_{0}\left( -t\right) V\left(
s\right) U_{0}\left( t\right) V\left( t\right) $ for all $t,s\geq 0$.

Conversely, given a strongly continuous group $U_{0}$ and a strongly
continuous $U_{0}$-cocycle $V$, then we may define a family $U$ by 
\begin{equation*}
U\left( t\right) =\left\{ 
\begin{array}{cc}
U_{0}\left( t\right) V\left( t\right) , & t\geq 0; \\ 
V\left( -t\right) ^{\dag }U_{0}\left( t\right) , & t<0.
\end{array}
\right.
\end{equation*}
and this is readily seen to be a strongly continuous group. Consequently we
deduce the existence of an associated Hamiltonian $K$.

The somewhat surprising feature of the converse is that $V$ does not need to
be strongly differentiable. In particular, it applies to the class of
quantum stochastic evolutions modelling open system dynamics. In such cases,
the existence of Hamiltonians $K_{0}$ and $K$ is immediately apparent,
however, it must of necessity be the case that domains do not have dense
intersection and so the formal subtraction $\Upsilon =K-K_{0}$ does not lead
to a self-adjoint operator in any meaningful way.

A well studied model is the dilations of an irreversible dynamics using Fock
space over $L^{2}$-functions of time where the free dynamics is second
quantization of the time shift. It has been a long standing problem to
characterize the associated Hamiltonian $K$\ for these models \cite{Acc90}.
The major breakthrough came in 1997 when A.N. Chebotarev solved this problem
for the class of quantum stochastic evolutions satisfying
Hudson-Parthasarathy differential equations with bounded commuting system
coefficients \cite{Cheb97},\cite{Cheb98}. His insight was based on
scattering theory of a one-dimensional system with a Dirac potential, say,
with formal Hamiltonian 
\begin{equation*}
k=i\partial +E\delta
\end{equation*}
describing a one-dimensional particle propagating along the negative $x$%
-axis with a delta potential of strength $E$ at the origin. (In Chebotarev's
analysis the $\delta $-function is approximated by a sequence of regular
functions, and a strong resolvent limit is performed.) \ The mathematical
techniques used in this approach were subsequently generalized by Gregoratti 
\cite{Gregoratti} to relax the commutativity condition. More recently, the
analysis has been further extended to treat unbounded coefficients \cite{QG}.

Independently, several authors have been engaged in the program of
describing the Hamiltonian nature of quantum stochastic evolutions by
interpreting the time-dependent function $\Upsilon \left( t\right) $ as
being an expression involving quantum white noises satisfying a singular CCR 
\cite{Go97}\cite{Go97a}\cite{Go99}\cite{AVL}. This would naturally suggest
that $\Upsilon $ should be interpreted as a sesquilinear expression in these
noises at time $t=0$. Specifically one considered time dependent Wick
ordered expressions $\Upsilon \left( t\right) =\sum_{ij}E_{ij}a_{i}^{\dag
}\left( t\right) a_{j}\left( t\right) +\sum_{i}E_{i0}a_{i}^{\dag }\left(
t\right) +\sum_{j}E_{0j}a_{j}\left( t\right) +E_{00}$ where the $%
E_{ij}=E_{ji}^{\dag }$ are operators on the initial space and $a_{i}\left(
t\right) ,a_{i}^{\dag }\left( t\right) $ are delta-correlated quantum white
noises corresponding to the formal derivatives of the annihilation and
creation. process. With $a_{i}\left( t\right) =U_{0}\left( -t\right)
a_{i}\left( 0\right) U_{0}\left( t\right) $, we see that $\Upsilon \left(
t\right) \equiv U_{0}\left( -t\right) \Upsilon U_{0}\left( t\right) $ where $%
\Upsilon =\Upsilon \left( 0\right) $. We shall show that this intuition
essentially provides the correct answer, though without using quantum white
noise! This approach inspired W. von Waldenfels to give an alternative
construction of the associated Hamiltonian, for diffusions \cite{WvW05} and
simple jump processes \cite{WvW05a}, however this was formulated through the
conventions of kernel calculus.

The aim of the current paper is to complete this program by returning to the
original one-dimensional model considered by Chebotarev. We alternatively
consider this as a problem of finding a suitable self-adjoint for the
singular Hamiltonian \cite{RSII}. Here the generator of the free dynamics $%
k_{0}=i\partial $ is not semi-bounded and the $\delta $-perturbation is
viewed as a singular rank-one perturbation. We employ methods introduced by
Albeverio and Kurasov \cite{AlKur97}\cite{AlKur99a} \cite{AlKurBook00} to
construct self-adjoint extensions of such models. In particular, the
boundary conditions that ones imposes at the origin corresponds to a phase
change of $s=\frac{1-\frac{i}{2}E}{1+\frac{i}{2}E}$, which should be
contrasted with the condition $s=e^{-iE}$ deduced by Chebotarev. We show
that the constructions of Albeverio and Kurasov are amenable to second
quantization and the form of $\Upsilon $ suggested by quantum white noise
analysis is precisely what is needed to obtain the description of the
associated Hamiltonian $K$ derived by Chebotarev, Gregoratti and von
Waldenfels. The construction avoids both the use of quantum white noises and
the unwieldy complexity of the kernel calculus by instead using the defect
vectors which lie in the underlying one-particle Hilbert space. One
important subtly is that the continuous tensor product decomposition for
Fock spaces is here implemented by the Sobolev space $W^{1,2}$ inner product
and not the usual $L^{2}$ inner product.

We show that there is a natural class of boundary conditions giving rise to
different extensions, and therefore different physical representations of
the same problem. These are parameterized by what effectively are the
additional complex damping terms (describing for instance energy shifts,
e.g., the Lamb shift \cite{AGL}) and which have earlier been referred to as
a degree of ``gauge freedom''.

\section{Singular Perturbations}

We shall adopt the standard convention that the sesquilinear Hilbert space
inner product is conjugate linear in the first argument and linear in the
second. Let $\xi $ be a conjugate linear functional on some domain of
functions on the real line. Its adjoint $\xi ^{\dag }$ is the linear
functional on the same domain defined through complex conjugation, that is, $%
\xi ^{\dag }\left( \phi \right) =\xi \left( \phi \right) ^{\ast }$. With a
standard abuse of the Dirac bra-ket notation, we shall write $\xi ^{\dag
}\left( \phi \right) =\langle \xi |\phi \rangle ,\xi \left( \phi \right)
=\langle \phi |\xi \rangle $, or more simply 
\begin{equation*}
\xi =|\xi \rangle ,\quad \xi ^{\dag }=\langle \xi |,
\end{equation*}
even though the functionals need not correspond to vectors in the Hilbert
space $L^{2}\left( \mathbb{R}\right) $.

The weak derivative of a measurable function $\phi $ will be denoted as $%
\partial \phi $. For a fixed $\frak{H}$\ be a Hilbert space and $I$ an open
subset of $\mathbb{R}^{d}$, we shall write $W^{1,2}\left( I,\frak{H}\right) $
for the Sobolev space of $\frak{H}$-valued functions $\phi $ possessing a
(weak) derivative and such that both such that $\phi $ and $\partial \phi $
are square-integrable. This is again a Hilbert space with inner product 
\begin{equation*}
\left\langle \phi |\psi \right\rangle _{1,2}\triangleq \int_{I}\left( \phi
^{\ast }\psi +\partial \phi ^{\ast }\partial \psi \right) .
\end{equation*}
Note that the corresponding norm $\left\| \cdot \right\| _{1,2}$ is then the
graph norm associated with the first derivative operator $\partial $.

Formally one may consider the Hamiltonian 
\begin{equation*}
k=i\partial +E\delta
\end{equation*}
describing a one-dimensional particle propagating along the negative $x$%
-axis with a delta potential of strength $E$ at the origin. As a
mathematical problem, we then have a perturbation of a self-adjoint operator 
$k_{0}=i\partial $, which is not semi-bounded, by a rank-one perturbation $%
\Upsilon =E|\delta \rangle \langle \delta |$ - that is to say $\Upsilon \phi
\left( x\right) =\phi \left( 0\right) \delta \left( x\right) $. The Dirac
functional $\delta $ is however bounded on the domain of $k_{0}$. The
singular nature of the potential implies that the particle wavefunction will
be discontinuous at the origin. In particular, such functions will not be in
the domain of the operator $k_{0}$.

We now review the theory of self-adjoint extensions of the generator of
linear translations $i\partial $ on the punctured line $\mathbb{R}/\left\{
0\right\} $.

\subsection{Distributions on discontinuous test functions}

Let $AC\left( I\right) $ denote the set of absolutely continuous functions
on an open subset $I$ of the real line. We consider some singular
functionals on this space.

\begin{definition}
The one-sided delta functionals $\delta _{\pm }$ on $AC\left( \mathbb{R}%
/\left\{ 0\right\} \right) $\ are given by 
\begin{equation*}
\left\langle \delta _{\pm }|\phi \right\rangle =\phi \left( 0^{\pm }\right) ,
\end{equation*}
and we introduce the associated functionals: 
\begin{equation*}
\begin{array}{rr}
\text{(jump at the origin)} & \left\langle \jmath \right| \triangleq
\left\langle \delta _{+}-\delta _{-}\right| , \\ 
\text{(symmetric delta functional)} & \left\langle \delta _{\star }\right|
\triangleq \frac{1}{2}\left\langle \delta _{+}+\delta _{-}\right| .
\end{array}
\end{equation*}
That is, $\langle \jmath |\phi \rangle =\phi \left( 0^{+}\right) -\phi
\left( 0^{-}\right) $ and $\langle \delta _{\star }|\phi \rangle =\frac{1}{2}%
\phi \left( 0^{+}\right) +\frac{1}{2}\phi \left( 0^{-}\right) $.
\end{definition}

Note that we have 
\begin{equation*}
|\delta _{\pm }\rangle =|\delta _{\ast }\rangle \pm \frac{1}{2}|\jmath
\rangle .
\end{equation*}

\begin{lemma}
Introducing the form $\mathcal{J}\triangleq |\delta _{+}\rangle \left\langle
\delta _{+}\right| -|\delta _{-}\rangle \left\langle \delta _{-}\right| $on
domain $AC\left( \mathbb{R}/\left\{ 0\right\} \right) $, we have the
following identity 
\begin{equation}
\mathcal{J}=|\jmath \rangle \left\langle \delta _{\star }\right| +|\delta
_{\star }\rangle \left\langle \jmath \right| .  \label{jumps}
\end{equation}
\end{lemma}

\begin{proof}
The right hand side is $\frac{1}{2}|\delta _{+}-\delta _{-}\rangle
\left\langle \delta _{+}+\delta _{-}\right| +\frac{1}{2}|\delta _{+}+\delta
_{-}\rangle \left\langle \delta _{+}-\delta _{-}\right| $ and expanding out
gives the result.
\end{proof}

\subsection{Distributional First Order Derivatives}

The differential operator $k_{0}\equiv i\partial $ is symmetric on\thinspace 
$W^{1,2}\left( \mathbb{R}\right) $ and its closure is the generator of
translations of wavefunctions along the real axis. This is not true when we
try restrict to\thinspace $W^{1,2}\left( \mathbb{R}/\left\{ 0\right\}
\right) $ due to the jump discontinuity at the origin \ In fact, $i\partial $
is not now symmetric as we have the integration by parts formula 
\begin{equation*}
\left\langle \psi |\partial \phi \right\rangle +\left\langle \partial \psi
|\phi \right\rangle =\left. \psi ^{\ast }\phi \right| _{0^{-}}^{0^{+}}\equiv
\left\langle \psi |\mathcal{J}\phi \right\rangle ,
\end{equation*}
for $\phi ,\psi \in W^{1,2}\left( \mathbb{R}/\left\{ 0\right\} \right) $.
Instead, let us introduce the distributional derivative operator 
\begin{equation}
iD\triangleq i\partial +i|\delta _{\star }\rangle \left\langle \jmath
\right| .  \label{iD}
\end{equation}
A combination of the lemma and the integration-by-parts formula shows that $%
iD$ is symmetric operator on $W^{1,2}\left( \mathbb{R}/\left\{ 0\right\}
\right) $. Note that $D\phi $ is then typically a functional on $%
W^{1,2}\left( \mathbb{R}/\left\{ 0\right\} \right) $ and not a function.

\subsection{Albeverio-Kurasov Construction}

Albeverio and Kurasov study rank-one perturbations of self adjoint operators 
$k_{0}$ that are not semibounded, specifically they study formal of the form 
$k_{0}+|\varphi \rangle \left\langle \varphi \right| $ where $\left\langle
\varphi \right| $ is a bounded functional on $dom(k_{0})$ though the
perturbation need not be form bounded. The consider the restriction $\tilde{k%
}_{0}$ of $k_{0}$ to the dense domain $dom\tilde{(}k_{0})=\left\{ \psi \in
dom(k_{0}):\left\langle \varphi \right| \psi \rangle =0\right\} $, and study
the scales of Hilbert spaces associated with the adjoint $\tilde{k}%
_{0}^{\dag }$.

In our case, $k_{0}=i\partial $ with domain $W^{1,2}\left( \mathbb{R}\right) 
$ and, since we consider $\delta $-function perturbations, the restricted
domain will be taken to be $V_{0}=\{\psi \in W^{1,2}\left( \mathbb{R}\right)
:\psi \left( 0\right) =0\}$. In this $\tilde{k}_{0}^{\dag }$ will be the
operator $i\partial $ with domain $W^{1,2}\left( \mathbb{R}/\left\{
0\right\} \right) $.

\subsection{Deficiency Subspaces}

Let us introduce the following pair of vectors $\phi _{\pm }\in
W^{1,2}\left( \mathbb{R}/\left\{ 0\right\} \right) $: 
\begin{equation}
\phi _{\pm }\left( t\right) =\mp ie^{\pm t}1_{\left( 0,\infty \right)
}\left( \pm t\right) .
\end{equation}
An elementary calculation shows that $\partial \left| \phi _{\pm
}\right\rangle =\mp \left| \phi _{\pm }\right\rangle $ and that $%
\left\langle \jmath |\phi _{\pm }\right\rangle =-i$, so that $iD\left| \phi
_{\pm }\right\rangle =\mp i\left| \phi _{\pm }\right\rangle +\left| \delta
_{\ast }\right\rangle $, or 
\begin{equation*}
\left| \phi _{\pm }\right\rangle =\left( iD\pm i\right) ^{-1}\left| \delta
_{\ast }\right\rangle .
\end{equation*}
More generally $iD_{\sigma }\left| \phi _{\pm }\right\rangle =\mp i\left|
\phi _{\pm }\right\rangle +\left| \zeta \right\rangle $.

It is convenient to realize $iD$ as being the adjoint $\tilde{k}_{0}^{\dag }$
to the operator $\tilde{k}_{0}$ as the restriction of $i\partial $ to the
dense domain $V_{0}=\left\{ \psi \in W^{1,2}\left( \mathbb{R}\right) :\psi
\left( 0\right) =0\right\} $. Note that $V_{0}$ is a Hilbert space with the
Sobolev inner product. The deficiency subspaces $V_{\pm }=\ker (\tilde{k}%
_{0}^{\dag }\pm i)$ are then both one-dimensional and spanned by the\ defect
vectors $\phi _{\pm }$ respectively: $V_{\pm }=\left\{ \mathbb{C}\phi _{\pm
}\right\} $. The domain of the adjoint can then be written as 
\begin{equation}
dom(\tilde{k}_{0}^{\dag })=V_{0}\oplus _{1,2}V_{+}\oplus _{1,2}V_{-}
\label{direct sum}
\end{equation}
where the three subspaces are mutually orthogonal with respect to the
Sobolev space inner product. (See, for instance, theorem X.2 \cite{RSII}.)

The elements of $dom(\tilde{k}_{0}^{\dag })$ can then be represented as 
\begin{equation*}
\left| \psi \right\rangle =\left| \psi _{0}\right\rangle +c_{+}\left| \phi
_{+}\right\rangle +c_{-}\left| \phi _{-}\right\rangle
\end{equation*}
where $\psi _{0}\in V_{0}$ and $c_{\pm }=\pm i\psi \left( 0^{\pm }\right)
\in \mathbb{C}$, the action of the adjoint is 
\begin{equation*}
\tilde{k}_{0}^{\dag }\left| \psi \right\rangle =i\partial \left| \psi
_{0}\right\rangle -ic_{+}\left| \phi _{+}\right\rangle +ic_{-}\left| \phi
_{-}\right\rangle .
\end{equation*}
We therefore represent $iD$ by $\tilde{k}_{0}^{\dag }$ and to check
consistency, we note that for $\psi \in dom(\tilde{k}_{0}^{\dag })$%
\begin{eqnarray*}
iD\left| \psi \right\rangle &=&i\partial \left| \psi _{0}\right\rangle
+i\psi \left( 0^{+}\right) \left( -i\left| \phi _{+}\right\rangle +\left|
\delta _{\ast }\right\rangle \right) -i\psi \left( 0^{-}\right) \left(
i\left| \phi _{-}\right\rangle +\left| \delta _{\ast }\right\rangle \right)
\\
&=&i\partial \left| \psi \right\rangle +i\left\langle \jmath |\psi
\right\rangle \left| \delta _{\ast }\right\rangle .
\end{eqnarray*}
For $\phi ,\psi \in dom(\tilde{k}_{0}^{\dag })$, the associated boundary
form $\left[ \phi |\psi \right] :=\langle \phi |\tilde{k}_{0}^{\dag }\psi
\rangle +\langle \tilde{k}_{0}^{\dag }\phi |\psi \rangle $ will be $\left[
\phi |\psi \right] =-i\left\langle \phi |\mathcal{J}\psi \right\rangle $.
All self-adjoint extensions of $k_{0}$\ can then be parameterized by a
unimodular complex parameter $s$. We define $k_{0,s}$ as the restriction of $%
\tilde{k}_{0}^{\dag }$ to the domain 
\begin{equation*}
dom\left( \tilde{k}_{0,s}\right) :=\left\{ \psi \in dom(\tilde{k}_{0}^{\dag
}):\psi \left( 0^{+}\right) =s\psi \left( 0^{-}\right) \right\} ,
\end{equation*}
where the functions have an abrupt phase change $s$ as we pass through the
origin.

\subsection{Singular Perturbations of $i\partial $}

We return to the formal Hamiltonian $k=i\partial +E\delta $ which we now
interpret as 
\begin{equation*}
k=iD+E\left| \delta _{\ast }\right\rangle \left\langle \delta _{\ast }\right|
\end{equation*}
The Hamiltonian can then be split up into continuous and singular components 
$k=k_{\text{ac}}+k_{\text{sing}}$ where $k_{\text{ac}}=i\partial $ and 
\begin{equation*}
k_{\text{sing}}=i\left| \delta _{\ast }\right\rangle \langle \jmath
|+E\left| \delta _{\ast }\right\rangle \left\langle \delta _{\ast }\right| .
\end{equation*}
Generally speaking, $k\left| \psi \right\rangle $ will be a functional for
given $\psi \in dom(\tilde{k}_{0}^{\dag })$ except however when $k_{\text{%
sing}}\left| \psi \right\rangle =0$ and the space of vector for which this
vanishes defines the domain of $k$. This requires that $\left\{ i\langle
\jmath |+E\left\langle \delta _{\ast }\right| \right\} \left| \psi
\right\rangle =0$, or 
\begin{equation*}
i\left[ \psi \left( 0^{+}\right) -\psi \left( 0^{-}\right) \right] +\frac{1}{%
2}E\left[ \psi \left( 0^{+}\right) +\psi \left( 0^{-}\right) \right] =0.
\end{equation*}
This can be written as the boundary condition 
\begin{equation}
\psi \left( 0^{-}\right) =\dfrac{1-\frac{i}{2}E}{1+\frac{i}{2}E}\psi \left(
0^{+}\right) .  \label{first bc}
\end{equation}
The domain of $k$ is therefore the set of $\psi \in W^{1,2}\left( \mathbb{R}%
/\left\{ 0\right\} \right) $ satisfying $\left( \ref{first bc}\right) $.
Note that the phase change $s=\dfrac{1-\frac{i}{2}E}{1+\frac{i}{2}E}$ is the
Cayley transform of $\frac{1}{2}E.$

\section{Second Quantization}

We now consider the Hilbert space 
\begin{equation*}
\frak{H}=\frak{h}\otimes \Gamma \left( L_{\frak{K}}^{2}\left( \mathbb{R}%
,dt\right) \right)
\end{equation*}
where $\frak{h}$,$\frak{K}$ are fixed separable Hilbert spaces and $\Gamma
\left( \cdot \right) $ is the bosonic Fock functor. A typical vector in $%
\frak{H}$ is $\Phi =\left( \Phi _{n}\right) $ where, for $n=0,1,2,\cdots ,$
we have that $\Phi _{n}$ is a $\frak{h}\otimes \frak{K}^{\otimes n}$-valued
function on $\mathbb{R}^{n}$ symmetric in all its arguments and such that 
\begin{equation*}
\sum_{n\geq 0}\frac{1}{n!}\int_{\mathbb{R}^{n}}\left\| \Phi _{n}\left(
t_{1},\cdots ,t_{n}\right) \right\| ^{2}dt_{1}\cdots dt_{n}<\infty .
\end{equation*}

We shall treat the case $\frak{K}=\mathbb{C}$ initially for transparency.

\subsection{Second quantization of $iD$}

Fock space has the continuous tensor product decomposition $\Gamma \left( 
\frak{h}_{1}\oplus \frak{h}_{2}\right) \cong \Gamma \left( \frak{h}%
_{1}\right) \oplus \Gamma \left( \frak{h}_{2}\right) $ and we exploit the
direct sum decomposition $\left( \ref{direct sum}\right) $ to write 
\begin{equation}
\Gamma \left( dom(k_{0}^{\dag })\right) \cong \Gamma _{0}\otimes
_{1,2}\Gamma _{+}\otimes _{1,2}\Gamma _{-},  \label{Fock decomp}
\end{equation}
where $\Gamma _{\#}=\Gamma \left( V_{\#}\right) $ for $\#=0,+,-$, and the
tensor product is with respect to the Sobolev inner product. Recall that
every $\psi \in dom(\tilde{k}_{0}^{\dag })$ will have the decomposition $%
\psi =\psi _{0}+i\psi \left( 0^{+}\right) \phi _{+}-i\psi \left(
0^{-}\right) \phi _{-}$ and can define the usual exponential vectors $%
\varepsilon \left( \psi \right) =\sum_{n\geq 0}\frac{1}{n!}\otimes
_{1,2}^{n}\psi $ and on this domain define the annihilation fields $A\left(
\phi \right) $, with $\phi \in dom(\tilde{k}_{0}^{\dag })$, by 
\begin{equation*}
A\left( \phi \right) \varepsilon \left( \psi \right) =\left\langle \phi
|\psi \right\rangle _{1,2}\varepsilon \left( \psi \right) \text{.}
\end{equation*}
where $\left\langle \phi |\psi \right\rangle _{1,2}\equiv \left\langle \phi
_{0}|\psi _{0}\right\rangle _{1,2}+\psi ^{\ast }\left( 0^{+}\right) \phi
\left( 0^{+}\right) +\psi ^{\ast }\left( 0^{-}\right) \phi \left(
0^{-}\right) $.

Now the spaces $V_{\pm }$ are both one-dimensional and so the Fock spaces $%
\Gamma \left( V_{\pm }\right) $ each correspond to the Hilbert space of an
independent single mode harmonic oscillator and we take the respective
annihilator operators to be 
\begin{equation}
a_{\pm }\triangleq A\left( \pm i\phi _{\pm }\right) .
\end{equation}
With this convention, we have for $\phi \in dom(K_{0}^{\dag })$, $A\left(
\phi \right) =A(\phi _{0})+\phi \left( 0^{+}\right) a_{+}+\phi \left(
0^{-}\right) a_{-}$. In particular, since $\left\langle \pm i\phi _{\pm
}|\psi \right\rangle _{1,2}=\phi \left( 0^{\pm }\right) $, we have 
\begin{equation*}
a_{\pm }\varepsilon \left( \psi \right) =\psi \left( 0^{\pm }\right)
\,\varepsilon \left( \psi \right) .
\end{equation*}
It is convenient to introduce 
\begin{equation*}
a_{\star }\triangleq \frac{1}{2}\left( a_{+}+a_{-}\right)
\end{equation*}

The second quantization of $\tilde{k}_{0}^{\dag }=iD$ is then given by $%
K_{0}=K_{0,\text{ac}}+K_{0,\text{sing}}$ where 
\begin{equation}
K_{0,\text{ac}}=d\Gamma \left( i\partial \right) ,\quad K_{0,\text{sing}%
}=id\Gamma \left( |\delta _{\star }\rangle \left\langle \jmath \right|
\right) =ia_{\star }^{\dag }\left( a_{+}-a_{-}\right) .
\end{equation}

\subsection{Singular Perturbations of $K_{\text{ac}}=d\Gamma \left(
i\partial \right) $}

We now consider the perturbed Hamiltonian $K=K_{0}+\Upsilon $ where 
\begin{equation}
\Upsilon =E_{11}a_{\star }^{\dag }a_{\star }+E_{10}a_{\star }^{\dag
}+E_{01}a_{\star }+E_{00}
\end{equation}
where the $E_{\alpha \beta }$ are operators on $\frak{h}$ with $\left(
E_{\alpha \beta }\right) ^{\dag }=E_{\beta \alpha }$. The Hamiltonian may
then be decomposed into continuous and singular parts as 
\begin{equation*}
K_{\text{ac}}+E_{00}+E_{01}a_{\star }+a_{\star }^{\dag }\left(
ia_{+}-ia_{-}+E_{11}a_{\star }+E_{10}\right) .
\end{equation*}
Let us introduce the following operators on the system space 
\begin{equation}
S=\frac{1-\frac{i}{2}E_{11}}{1+\frac{i}{2}E_{11}},\quad L=-\frac{i}{1+\frac{i%
}{2}E_{11}}E_{10},\quad H=E_{00}+\frac{1}{2}E_{01}\func{Im}\left\{ \frac{1}{%
1+\frac{i}{2}E_{11}}\right\} E_{10}.  \label{HP parameters}
\end{equation}

\begin{theorem}
The domain of the self-adjoint extension corresponding to the operator $%
K=K_{0}+\Upsilon $ is the set of vectors $\Phi \in \frak{h}\otimes
dom(K_{0}) $ satisfying the boundary conditions 
\begin{equation}
a_{-}\Phi =Sa_{+}\Phi +L,  \label{2nd bc}
\end{equation}
and on this domain we have 
\begin{equation}
-iH_{\text{total}}\Phi =\left( -iK_{\text{ac}}-\frac{1}{2}L^{\dag
}L-iH\right) \Phi -L^{\dag }Sa_{+}\Phi .
\end{equation}
\end{theorem}

\begin{proof}
Our strategy is basically the same as in the one-particle case - we choose
the domain of the operator to consist of those vectors $\Phi $ such that $%
\left( ia_{+}-ia_{-}+E_{11}a_{\star }+E_{10}\right) \Phi =0$. This is
equivalent to the boundary condition $\left( \ref{2nd bc}\right) $. For
vectors on this domain, we may substitute $a_{-}\Phi $ for $a_{+}\Phi $
using the boundary condition and this gives the desired result.
\end{proof}

\subsection{Multiple Field Modes}

We now consider the more general case where $\frak{K}=\mathbb{C}^{n}$. Let $%
\left\{ e_{j}:j=1,\cdots ,n\right\} $ be an orthonormal basis for $\frak{K}$%
. The one-particle space is now $L_{\mathbb{C}^{n}}^{2}\left( \mathbb{R}%
,dt\right) =\mathbb{C}^{n}\otimes L^{2}\left( \mathbb{R},dt\right) $ and we
replace the decomposition $\left( \ref{Fock decomp}\right) $ by $\Gamma
\left( \mathbb{C}^{n}\otimes dom(\tilde{k}_{0}^{\dag })\right) =\Gamma
_{0}^{\left( n\right) }\otimes _{1,2}\Gamma _{+}^{\left( n\right) }\otimes
_{1,2}\Gamma _{-}^{\left( n\right) }$ where $\Gamma _{\#}^{\left( n\right)
}=\Gamma \left( \mathbb{C}^{n}\otimes V_{\#}\right) $. The defect vectors
for the ampliation of $\tilde{k}_{0}$ to $\mathbb{C}^{n}\otimes L^{2}\left( 
\mathbb{R},dt\right) $ may now be fixed as $e_{j}\otimes \phi _{\pm }$ and
so the deficiency indices are now $\left( n,n\right) $.

Proceeding as before, we introduce the independent annihilators 
\begin{equation*}
a_{j,\pm }\triangleq A\left( \pm ie_{j}\otimes \phi _{\pm }\right)
\end{equation*}
and the second quantization of the ampliation of $\tilde{k}_{0}^{\dag }$ is $%
K_{0}=K_{0,\text{ac}}+K_{0,\text{sing}}$ with 
\begin{equation}
K_{0,\text{ac}}=d\Gamma \left( i\partial \right) ,\quad K_{0,\text{sing}%
}=ia_{j,\star }^{\dag }\left( a_{j,+}-a_{j,-}\right) .
\end{equation}
(A summation over the range $1,\cdots ,n$ is implied for repeated Latin
indices!)

Let us also set $a_{0,\pm }\equiv 1$. As perturbation we consider the
singular term 
\begin{equation}
\Upsilon =E_{\alpha \beta }\,a_{\alpha }^{\dag }a_{\beta }.  \label{Upsilon}
\end{equation}
(Here we understand repeated Greek indices as implying a sum over $%
0,1,\cdots ,n$.) Again the $\left( n+1\right) ^{2}$ operators $E_{\alpha
\beta }$ are operators on $\frak{B}\left( \frak{h}\right) $ and we require \
that $\left( E_{\alpha \beta }\right) ^{\dag }=E_{\beta \alpha }$. It is
convenient to assemble them into a matrix $\mathbf{E}$. More generally we
consider the class of matrices 
\begin{equation*}
\mathbf{X}=\left( 
\begin{tabular}{ll}
$X_{00}$ & $X_{0\ell }$ \\ 
$X_{\ell 0}$ & $X_{\ell \ell }$%
\end{tabular}
\right) \in \frak{B}\left( \left( \mathbb{C}\oplus \frak{K}\right) \otimes 
\frak{h}\right)
\end{equation*}
where $X_{00}\in \frak{B}\left( \frak{h}\right) ,X_{0\ell }\in \frak{B}%
\left( \frak{h},\frak{K}\otimes \frak{h}\right) ,X_{\ell 0}\in \frak{B}%
\left( \frak{K}\otimes \frak{h},\frak{h}\right) $ and $X_{\ell \ell }\in 
\frak{B}\left( \frak{K}\otimes \frak{h}\right) .$ That is $X_{0\ell }$ is
the row vector $\left( X_{01},\cdots ,X_{0n}\right) $ and $X_{\ell 0}$ is
the column vector $\left( X_{01},\cdots ,X_{0n}\right) ^{T}$ while $X_{\ell
\ell }=\left( X_{ij}\right) $.

Let us also introduce the special matrix $\mathbf{\Pi }$ projecting onto the
subspace $\frak{K}\otimes \frak{h}$ of $\left( \mathbb{C}\oplus \frak{K}%
\right) \otimes \frak{h}\equiv \frak{h}\oplus \left( \frak{K}\otimes \frak{h}%
\right) $, that is, 
\begin{equation*}
\mathbf{\Pi }=\left( 
\begin{array}{cc}
0 & 0 \\ 
0 & 1
\end{array}
\right) .
\end{equation*}
Its coefficients are the Evans-Hudson delta 
\begin{equation*}
\hat{\delta}_{\alpha \beta }=\left\{ 
\begin{array}{cc}
1, & \alpha =\beta \in \left\{ 1,\cdots ,n\right\} , \\ 
0, & \text{otherwise.}
\end{array}
\right.
\end{equation*}

Given a self-adjoint $\frak{B}\left( \frak{h}\right) $-valued matrix $%
\mathbf{E}$\ as above, we recall some related matrices, originally
introduced in \cite{Go06}, starting with the matrix $\mathbf{G}$ defined by
the identity $\mathbf{G}=-i\mathbf{E-}\frac{i}{2}\mathbf{E\Pi G}$, 
\begin{eqnarray*}
\text{(It\={o}) }\mathbf{G} &\triangleq &-i\left( \mathbf{1}+\frac{i}{2}%
\mathbf{E\Pi }\right) ^{-1}\mathbf{E}=\left( 
\begin{tabular}{rr}
$-\frac{1}{2}L^{\dag }L-iH$ & $-L^{\dag }S$ \\ 
$L$ & $S-1$%
\end{tabular}
\right) , \\
\text{(Model) }\mathbf{V} &\triangleq &\mathbf{G}+\mathbf{\Pi }=\left( 
\begin{tabular}{rr}
$-\frac{1}{2}L^{\dag }L-iH$ & $-L^{\dag }S$ \\ 
$L$ & $S$%
\end{tabular}
\right) , \\
\text{(Galilean) }\mathbf{M} &\triangleq &\left( \mathbf{1}+\mathbf{\Pi G}%
\right) =\left( 
\begin{tabular}{rr}
$1$ & $0$ \\ 
$L$ & $S$%
\end{tabular}
\right) , \\
\text{(Dressing) }\mathbf{F} &\triangleq &\left( \mathbf{1}+\frac{1}{2}%
\mathbf{\Pi G}\right) =\left( \mathbf{1}+\frac{i}{2}\mathbf{\Pi E}\right)
^{-1}=\left( 
\begin{tabular}{rr}
$1$ & $0$ \\ 
$\frac{1}{2}L$ & $\frac{1}{2}\left( S+1\right) $%
\end{tabular}
\right) .
\end{eqnarray*}
Note the identity 
\begin{equation}
\mathbf{G}=-i\mathbf{EF}
\end{equation}
and that we encounter the operators 
\begin{eqnarray}
S &=&\left( 1-\frac{i}{2}E_{\ell \ell }\right) \left( 1+\frac{i}{2}E_{\ell
\ell }\right) ^{-1}\in \frak{B}\left( \frak{K}\otimes \frak{h}\right) , 
\notag \\
L &=&-i\left( 1+\frac{i}{2}E_{\ell \ell }\right) ^{-1}E_{\ell 0}\in \frak{B}%
\left( \frak{K}\otimes \frak{h},\frak{h}\right) ,  \notag \\
H &=&E_{00}+\frac{1}{2}\func{Im}E_{0\ell }\left( 1+\frac{i}{2}E_{\ell \ell
}\right) ^{-1}E_{\ell 0}.
\end{eqnarray}

With these conventions we may write 
\begin{equation*}
K_{\text{sing}}=i\mathbf{a}_{\star }^{\dag }\mathbf{\Pi }\left( \mathbf{a}%
_{+}-\mathbf{a}_{-}\right) ,\quad \Upsilon =\mathbf{a}_{\star }^{\dag }%
\mathbf{Ea}_{\star },
\end{equation*}
where 
\begin{equation*}
\mathbf{a}_{\pm }\triangleq \left( 
\begin{array}{c}
1 \\ 
a_{1,\pm } \\ 
\vdots \\ 
a_{n,\pm }
\end{array}
\right) ,\quad \mathbf{a}_{\star }=\frac{1}{2}(\mathbf{a}_{+}+\mathbf{a}%
_{-}).
\end{equation*}
The Hamiltonian is then 
\begin{equation*}
K=d\Gamma \left( i\partial \right) +\mathbf{a}_{\star }^{\dag }\left\{ i%
\mathbf{\Pi }\left( \mathbf{a}_{+}-\mathbf{a}_{-}\right) +\mathbf{Ea}_{\star
}\right\}
\end{equation*}
and the domain is the set of vectors $\Phi $ such that\ $\mathbf{\Pi }%
\left\{ i\left( \mathbf{a}_{+}-\mathbf{a}_{-}\right) +\mathbf{Ea}_{\star
}\right\} \Phi =0$. This boundary condition ensures that all the $a_{j,\star
}^{\dag }$ terms vanish $\left( j=1,\cdots ,n\right) $ and may be
reformulated as 
\begin{equation*}
\mathbf{\Pi }\left( \mathbf{E}-2i\mathbf{\Pi }\right) \mathbf{a}_{+}\Phi +%
\mathbf{\Pi }\left( \mathbf{E}+2i\mathbf{\Pi }\right) \mathbf{a}_{-}\Phi =0
\end{equation*}
or 
\begin{equation*}
a_{j,-}\Phi =S_{jk}a_{k,+}\Phi +L_{j}\Phi
\end{equation*}
with $S\equiv S_{jk}\otimes \left| e_{j}\right\rangle \left\langle
e_{k}\right| $ and $L\equiv L_{j}\otimes \left| e_{j}\right\rangle $. We
have trivially that $a_{0,-}=a_{0,+}$ and the we may include this in the
boundary condition to write 
\begin{equation}
\mathbf{a}_{+}\Phi \mathbf{=Ma}_{-}\Phi .
\end{equation}
Using the boundary condition, we may write 
\begin{eqnarray*}
K\Phi &=&K_{\text{ac}}\Phi +\frac{1}{2}E_{0\alpha }\left\{ a_{\alpha
,+}+a_{\alpha ,-}\right\} \Phi \\
&=&K_{\text{ac}}\Phi +\frac{1}{2}E_{0\alpha }\left\{ \delta _{\alpha \beta
}+M_{\alpha \beta }\right\} a_{\beta ,+}\Phi
\end{eqnarray*}
however $\frac{1}{2}\mathbf{E}\left( \mathbf{1}+\mathbf{M}\right) =\mathbf{E}%
\left( \mathbf{1}+\frac{1}{2}\mathbf{\Pi G}\right) =\mathbf{EF}=i\mathbf{G}$%
. With this we may write the\ action of the Hamiltonian on vectors
satisfying the boundary conditions as 
\begin{equation*}
K\Phi =K_{\text{ac}}\Phi +iG_{0\alpha }a_{\beta ,+}\Phi .
\end{equation*}

We may summarize our findings\ using the model matrix $\mathbf{V}$ in the
following theorem.

\begin{theorem}
The Hamiltonian associated to $K=K_{0}+\Upsilon $ with perturbation $\left( 
\ref{Upsilon}\right) $ is defined on the domain $\Gamma \left( dom(\tilde{k}%
_{0}^{\dag })\right) $ satisfying the boundary condition $\mathbf{a}_{+}\Phi 
\mathbf{=Va}_{-}\Phi $. Its action on this domain, and the boundary
condition, are given by 
\begin{eqnarray*}
-iK\Phi &=&\left( V_{00}+V_{0k}a_{k,+}-i\tilde{K}_{0}\right) \Phi =\left(
V_{0\beta }a_{\beta }\left( 0^{+}\right) -i\tilde{K}_{0}\right) \Phi , \\
a_{j,-}\Phi &=&V_{j0}\Phi +V_{jk}a_{k,+}\Phi =V_{j\beta }a_{\beta ,+}\Phi .
\end{eqnarray*}
\end{theorem}

\section{Gauge freedom}

A complex number $\kappa _{+}$ with strictly positive real part will be
referred to as a \textit{complex damping constant}. For convenience we shall
normalize complex damping constants as 
\begin{equation*}
\kappa _{\pm }=\frac{1}{2}\pm i\sigma ,
\end{equation*}
where $\sigma $ is real. For a fixed complex damping $\kappa $ we then
define a functional $\zeta =\zeta _{\kappa }$ by 
\begin{equation}
\left| \zeta \right\rangle =\kappa _{+}\left| \delta _{+}\right\rangle
+\kappa _{-}\left| \delta _{-}\right\rangle \equiv \gamma \left| \delta
_{\star }\right\rangle +i\sigma \left| \jmath \right\rangle .
\end{equation}
We then we have the following local identity generalizing $\left( \ref{jumps}%
\right) $

\begin{equation}
\mathcal{J}=|\jmath \rangle \left\langle \zeta \right| +|\zeta \rangle
\left\langle \jmath \right| ,
\end{equation}
as the\ $|\jmath \rangle \left\langle \jmath \right| $ terms cancel. This
allows us to construct a more general class of self-adjoint extensions of
the restriction of $i\partial $. Let $\sigma $ be the imaginary part of the
complex damping $\kappa _{+}$ and define 
\begin{equation*}
iD_{\sigma }\triangleq i\partial +i|\delta _{\star }\rangle \left\langle
\jmath \right| -\sigma |\jmath \rangle \left\langle \jmath \right| \equiv
i\partial +i|\zeta \rangle \left\langle \jmath \right|
\end{equation*}
with $\zeta $ as above. It follows that $iD_{\sigma }$ is likewise a
symmetric operator on $W^{1,2}\left( \mathbb{R}/\left\{ 0\right\} \right) $.

The formal Hamiltonian $k=i\partial +E\delta $ may be alternatively
interpreted as 
\begin{equation*}
k_{\sigma }=iD_{\sigma }+E\left| \zeta \right\rangle \left\langle \zeta
\right| ,
\end{equation*}
or $k_{\sigma }=i\partial +i\left| \zeta \right\rangle \langle \jmath
|+E\left| \zeta \right\rangle \left\langle \zeta \right| $. We may follow
the same argument as before and arrange for the singular component to vanish
by imposing the boundary conditions: this time the condition $\left( \ref
{first bc}\right) $ is modified to $\psi \left( 0^{-}\right) =s_{\sigma
}\psi \left( 0^{+}\right) $ where we now have the phase $s_{\sigma }=\dfrac{%
1-i\kappa _{-}E}{1+i\kappa _{+}E}$.

This can be lifted immediately to the second quantization. For $\frak{K}=%
\mathbb{C}$, we set $\frak{a}=\kappa _{-}a_{+}+\kappa _{+}a_{-}$. We make
the corresponding replacements: $K_{0,\text{sing}}\left( \sigma \right) =i%
\frak{a}^{\dag }\left( a_{+}-a_{-}\right) $ and $\Upsilon \left( \sigma
\right) =E_{11}\frak{a}^{\dag }\frak{a}+E_{10}\frak{a}^{\dag }+E_{01}\frak{a}%
+E_{00}$. The boundary conditions arise from requiring the $\frak{a}^{\dag }$
terms to vanish and after similar algebra to before we arrive at the
restatement of the first theorem with the modified operators 
\begin{eqnarray}
S\left( \sigma \right) &=&\frac{1-i\kappa _{-}E_{11}}{1+i\kappa _{+}E_{11}}%
,\quad L\left( \sigma \right) =-\frac{i}{1+i\kappa _{+}E_{11}}E_{10},  \notag
\\
H\left( \sigma \right) &=&E_{00}+\func{Im}\left\{ E_{01}\frac{\kappa _{+}}{%
1+i\kappa _{+}E_{11}}E_{10}\right\} .
\end{eqnarray}
Identical expression where obtained for the singular limit of finite time
correlated Bose field in \cite{Go05}, theorem 8.1 equation $\left(
8.11\right) $.

The multiple field mode version of this is to introduce in place of $\mathbf{%
a}_{\star }=\frac{1}{2}(\mathbf{a}_{+}+\mathbf{a}_{-})$ the vector 
\begin{equation*}
\frak{a}=(\frac{1}{2}\mathbf{1}+i\mathbf{Z})\mathbf{a}_{+}+(\frac{1}{2}%
\mathbf{1}-i\mathbf{Z)a}_{-}
\end{equation*}
where $\mathbf{Z}=\left( 
\begin{array}{cc}
0 & 0 \\ 
0 & Z_{\ell \ell }
\end{array}
\right) $ is a hermitean matrix with complex entries. We then obtain a
straightforward modification of the second theorem with new coefficients $%
\mathbf{G}\left( \mathbf{E},\mathbf{Z}\right) =-i\left( \mathbf{1}+i\mathbf{E%
}\left( \frac{1}{2}\mathbf{\Pi }+i\mathbf{Z}\right) \right) ^{-1}\mathbf{E}$%
. This general type was first introduced in \cite{Go06}.

In general, the parameters $\left( Z_{jk}\right) $ may be termed ``gauge
parameters'' and by choosing a different set of parameters we obtain a
different self-adjoint extension. This type freedom/ambiguity is well-known
and is ultimately a question of fixing the desired physical model and cannot
arise from purely mathematical arguments alone \cite{RSII}.

\begin{acknowledgement}
The author wishes to thank Matthew James and Oleg G. Smolyanov for fruitful
discussions. He is particularly grateful to Alexei Iantchenko for
introducing him to reference \cite{AlKur99a}.
\end{acknowledgement}

\end{document}